# Radiation damage and thermal shock response of carbon-fiber-reinforced materials to intense high-energy proton beams

N. Simos,[1,*] Z. Zhong,[1] S. Ghose,[1] H. G. Kirk,[1] L-P Trung,[1] K. T. McDonald,[4] Z. Kotsina,[5] P. Nocera,[6] R. Assmann,[2] S. Redaelli,[2] A. Bertarelli,[2] E. Quaranta,[2] A. Rossi,[2] R. Zwaska,[3] K. Ammigan,[3] P. Hurh,[3] and N. Mokhov[3]

[1]*Brookhaven National Laboratory (BNL), Upton, New York 11973, USA*
[2]*CERN, CH-1211 Geneva, Switzerland*
[3]*Fermi National Accelerator Laboratory, Batavia, Illinois 60510-5011, USA*
[4]*Joseph Henry Laboratories, Princeton University, Princeton, New Jersey 08544, USA*
[5]*Department of Solid State Physics, National and Kapodistrian University of Athens, Athens 10679, Greece*
[6]*Department of Physics, University of Rome, 00185 Rome, Italy*

A comprehensive study on the effects of energetic protons on carbon-fiber composites and compounds under consideration for use as low-Z pion production targets in future high-power accelerators and low-impedance collimating elements for intercepting TeV-level protons at the Large Hadron Collider has been undertaken addressing two key areas, namely, thermal shock absorption and resistance to irradiation damage. Carbon-fiber composites of various fiber weaves have been widely used in aerospace industries due to their unique combination of high temperature stability, low density, and high strength. The performance of carbon-carbon composites and compounds under intense proton beams and long-term irradiation have been studied in a series of experiments and compared with the performance of graphite. The 24-GeV proton beam experiments confirmed the inherent ability of a 3D C/C fiber composite to withstand a thermal shock. A series of irradiation damage campaigns explored the response of different C/C structures as a function of the proton fluence and irradiating environment. Radiolytic oxidation resulting from the interaction of oxygen molecules, the result of beam-induced radiolysis encountered during some of the irradiation campaigns, with carbon atoms during irradiation with the presence of a water coolant emerged as a dominant contributor to the observed structural integrity loss at proton fluences $\geq 5 \times 10^{20}$ p/cm$^2$. The carbon-fiber composites were shown to exhibit significant anisotropy in their dimensional stability driven by the fiber weave and the microstructural behavior of the fiber and carbon matrix accompanied by the presence of manufacturing porosity and defects. Carbon-fiber-reinforced molybdenum-graphite compounds (MoGRCF) selected for their impedance properties in the Large Hadron Collider beam collimation exhibited significant decrease in postirradiation load-displacement behavior even after low dose levels ($\sim 5 \times 10^{18}$ p cm$^{-2}$). In addition, the studied MoGRCF compound grade suffered a high degree of structural degradation while being irradiated in a vacuum after a fluence $\sim 5 \times 10^{20}$ p cm$^{-2}$. Finally, x-ray diffraction studies on irradiated C/C composites and a carbon-fiber-reinforced Mo-graphite compound revealed (a) low graphitization in the "as-received" 3D C/C and high graphitization in the MoGRCF compound, (b) irradiation-induced graphitization of the least crystallized phases in the carbon fibers of the 2D and 3D C/C composites, (c) increased interplanar distances along the c axis of the graphite crystal with increasing fluence, and (d) coalescence of interstitial clusters after irradiation forming new crystalline planes between basal planes and excellent agreement with fast neutron irradiation effects.

## I. INTRODUCTION

Planned next-generation, multimegawatt power accelerators will require high-performance and high reliability secondary particle production targets to generate intense neutrino beams or spallation fields. Towards this end, the understanding of the behavior of target materials under extreme states of long-term irradiation combined with thermal shock must greatly expand to remain in step with the order of magnitude increase in the demand. Primaryconcerns are (a) the anticipated accumulation of radiation damage in the materials of choice and the resulting changes in their physiomechanical properties and (b) the ability of materials to withstand an intense beam-induced thermal shock while their physical properties are continuously undergoing changes. In targetlike applications such as the collimating structures of the 7-TeV proton beam at the Large Hadron Collider (LHC), where the selected material intercepts the beam halo, extreme dimensional stability and resilience to structural degradation is required.

*Corresponding author.
simos@bnl.gov

C/C composites consisting primarily of a carbon matrix reinforced with carbon fibers in 1D, 2D, and 3D weave manner have enjoyed applications in a number of industries, such as aerospace owed to their excellent thermal shock resistance, high specific strength, and modulus of elasticity along preferred fiber orientations. Their superior thermal resistance and ability to maintain strength at elevated temperatures has introduced them to applications with severe thermal conditions such as the thermal shield for space shuttle reentry. Serious consideration of these composites for application in very-high-temperature reactors and fusion reactors has been given due to the added properties of high sublimation temperature (~3600 K), high thermal conductivity, and low neutron absorption cross section. Studies at elevated temperatures of C/C aiming to understand the microstructural evolution of their complex structure with and in the absence of irradiating environments have been conducted. The microstructural evolutions of 3D C/C composite materials irradiated by carbon ions at elevated temperatures were studied by Tsai and co-workers [1] concluding that the composite exhibits a lower degree of graphitization than nuclear graphite and that modest irradiation levels of $7 \times 10^{21}$ C$^+$/m$^2$ at temperatures of 600°C led to observable radiation damage in the form of distinct cracks and grain distortions in the matrix as well as a severing of fibers. The structural modifications of C/C composites under high temperature and ion irradiation (2 keV H of low flux $5 \times 10^{16}$ ions m$^{-2}$ s$^{-1}$) were addressed by Paulmier and co-workers [2]. Their study revealed that the microstructure of the "as-received" carbon-carbon is strongly influenced by the processing route and that 2D C/C structures exhibit increased disorder with elevated temperatures.

Irradiation-induced structure and property changes in C-C composites for use in tokamak plasma facing was studied by Burchell [3]. Results of two irradiation experiments of 1D, 2D, and 3D carbon-composite structures irradiated at the High Flux isotope Reactor were reported. The study results indicated that the 3D C/C structure behaves more isotropically than two-directional or unidirectional composites and that the thermal conductivity, while severely degraded by neutron irradiation, can partially recover through annealing. The effects of carbon-fiber orientation and graphitization on solid state bonding of a C/C composite to nickel were studied by Nishida [4]. Specifically, C/C composites with the two types of carbon-fiber orientations, which are heat-treated at two different temperatures, were used, and the influences of both fiber orientation and graphitization on the joining of C/C composites to metals were investigated. Load-displacement tests were conducted revealing that the longitudinal section of carbon fiber undergoes slip deformation due to crystallographic anisotropy. A direct comparison of load-deflection data with results from the present study will be made later in a subsequent section. In Ref. [5], composite materials that include C/C composites and SiC/SiC in extreme radiation and temperatures for application in next-generation high-temperature reactors are discussed. Relevant to the behavior of C/C composites are studies conducted and reporting on nuclear graphite focusing on the microstructural and physical property evolution as a function of the temperature and fluence [5–11].

To assess the effects of energetic protons on carbon-fiber composites and compounds, a multifaceted experimental study has been undertaken addressing two key areas, namely, thermal shock absorption and proton irradiation damage under different environments and fluence levels. Specifically, intense shock generation by 24-GeV tight proton beam pulses and its absorption by carbon-fiber structures and graphite were studied experimentally to confirm the inherent ability of a 3D C/C fiber composite to withstand and absorb shock [12–15]. Key data from the experiment are reported in this paper. In addition, a series of proton irradiation campaigns were launched exploring the damage induced on various C/C structures and compounds such as carbon-fiber-reinforced Mo-graphite as a function of the proton fluence and irradiating environment. The irradiation campaigns were augmented by studies of high-temperature annealing and characterization of the carbon-fiber composites and compounds.

Presented in this paper are the results of the 24-GeV beam tests and a postirradiation assessment of the interaction of energetic protons with C/C composites and carbon-fiber-reinforced Mo-graphite compound. Macroscopically observed changes in the structure and physical properties are augmented with a microscopic assessment of the crystal structure changes that occur using high-energy x rays.

## II. EXPERIMENTAL

### A. Characterization

Using high-temperature annealing, differential scanning calorimetry and scanning electron microscopy (SEM) at the BNL Center of Functional Nanomaterials, C/C composites and the metal matrix compound MoGRCF were characterized and their evolution with temperature was observed. Figure 1 depicts SEM micrographs of the 3D C/C composite and MoGRCF compound while revealing the presence of manufacturing defects and cracks resulting from the differential expansion in fibers and matrix. Figure 2 shows the evolution of the MoGRCF metal matrix compound with a temperature up to 965°C. Carbon fibers integrated into the metal matrix composite are clearly visible at room temperature, while at 400°C a disordered structure has developed followed by the crystallization of phases in the compound at higher temperatures.

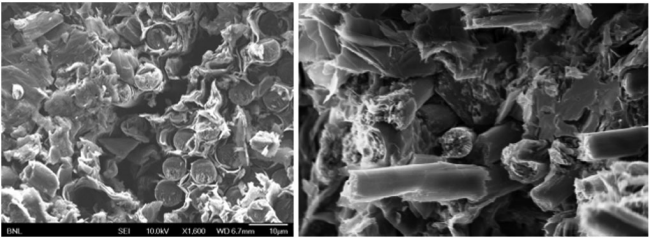

FIG. 1.   SEM micrographs of 3D C/C composite and MoGRCF compound at room temperature.

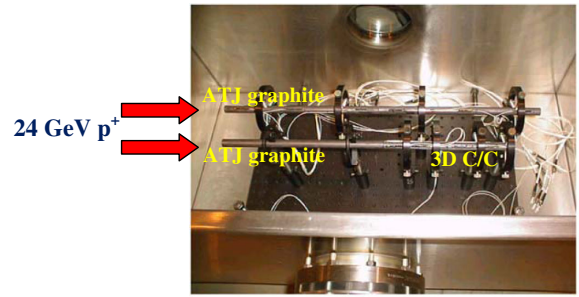

FIG. 3.   Setup of instrumented 3D C/C composite and ATJ graphite targets during the E951 24-GeV BNL AGS thermal shock experiments.

### B. Thermal shock experiments

Under the E-951 BNL experiment, 24-GeV proton tight beam pulses of intensity ~3.1 × 10¹² protons/pulse and $\sigma_x = 0.7$ mm, $\sigma_y = 1.7$ mm from the Accelerator Gradient Synchrotron (AGS) were directed on 3D C/C and ATJ grade graphite cylindrical targets of 1 cm radius and 12 cm length, respectively. The C/C and graphite rod targets shown in Fig. 3 were instrumented with Fabry-Perot fiber optic strain gauges capturing the dynamic response due to the induced thermal shock and comparing the response of the two carbon-based structures. The primary objectives of the experiment were the confirmation of the superior thermal shock resistance of C/C structures, as compared to graphite, and the generation of precise beam shock response data to facilitate the calibration of numerical models. The fiber optic strain gauges capturing the longitudinal responses were connected to a VELOCE signal capturing and processing system capable of resolving 500-kHz response signals. Two C/C cylindrical targets (12 cm each) were placed in tandem along the 24-GeV beam and downstream of a 24-cm-long instrumented graphite target.

The experiments revealed [12–15] that (a) the response of the 3D C/C is not one that would be expected by an amorphous structure (something that will be ideal for a high-power target design) and (b) the intensity of the response is decreased significantly as compared to the ATJ graphite, demonstrating the inherent ability of the composite to resist a thermally induced shock. Shown in Fig. 4 is a direct comparison of the captured microstrain response between graphite and 3D C/C. It is evident from the captured response that the C/C structure significantly decreases the shock-induced response as compared to graphite, thus confirming the superiority of composite structures in absorbing a shock.

### C. Irradiation damage experiments

Irradiation damage experiments were conducted at the Brookhaven Linear Isotope Producer (BLIP) beam line and in tandem operation with medical isotope production which typically utilizes the 118-MeV proton energy mode from the source and linac. The irradiation damage experiments of C/C composites and compounds used either the 200-MeV or the 181-MeV proton energies with the irradiation array upstream of the isotope-producing array. The irradiation array and the cooling between the specially designed capsules are precisely selected with the help of particle-tracking codes [16–18] such that the proton energy delivered to the isotope targets is identical to the energy profile of the 118-MeV operations. Figure 5 depicts the exact geometry and setup used in one of the irradiation campaigns to simulate energy loss and the desired beam profile (left) for a 200-MeV irradiation of MoGRCF and the actual target capsule array (right) lowered into the proton beam. Figure 5(a) depicts the code-estimated [17,18] proton profile through the entire array satisfying both the requirement that the energy over the isotope-producing array is undisturbed and that proton beam is fully arrested in the specially designed beam stop. Two beam current transformers along the BLIP beam line monitor the beam current to the targets. Beam position monitors as well as two plunging harps monitor the position and beam pulse shape. The latter is augmented with radiographs of thin nickel foils placed within the array [Fig. 7(c)].

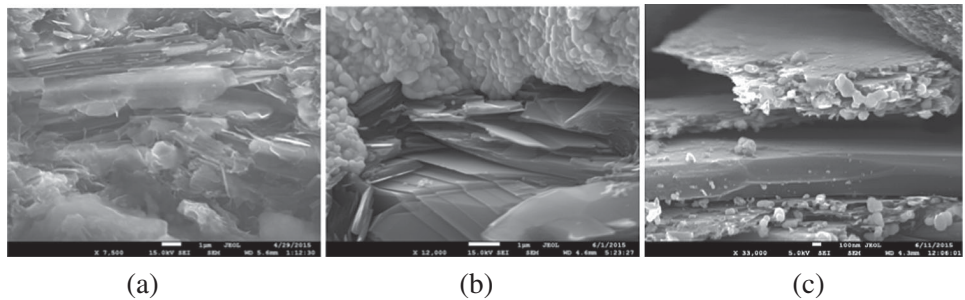

FIG. 2.   SEM micrographs of MoGRCF (atomic fraction: 21.51% Mo and 78.485% carbon) with integrated C/C fibers following annealing at elevated temperatures (a = 400°C, b = 720°C, c = 965°C).

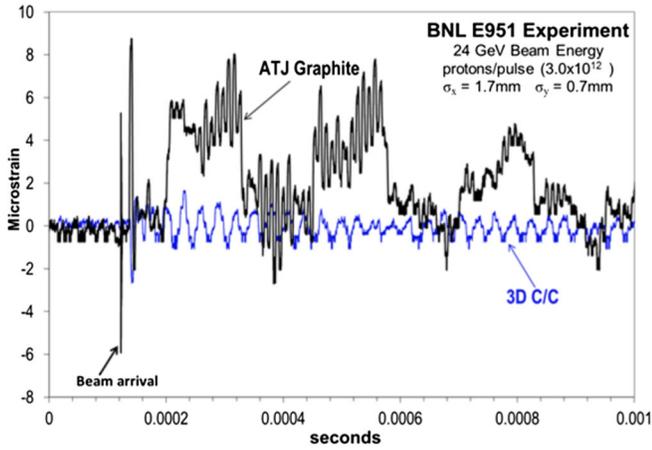

FIG. 4. Comparison of 3D C/C and graphite response subjected to the same intensity proton pulse.

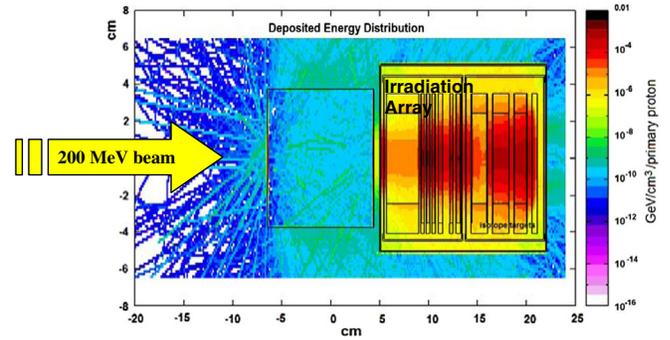

FIG. 6. Numerical modeling of the proton interaction with an irradiation target array in tandem with the isotope production array downstream.

Energy-deposition predictions using the Monte Carlo transport codes (MARS15 and FLUKA) throughout the irradiated array were made for all the irradiation campaigns or configurations and were used in subsequent thermomechanical analyses and experimental safety assessments. Figure 6 depicts energy-deposition results for a 200-MeV irradiation with irradiation target capsules in the upstream position and isotope targets (encapsulated RbCl) in the downstream position. The irradiation temperatures of the various composites being irradiated, whether encapsulated in the stainless steel capsules or directly cooled by the forced water flow across the cooling gaps, were estimated via a multifaceted numerical analysis and the predicted energy deposition. Specifically, a computational fluid dynamic (CFD) model was developed to calculate the heat transfer coefficients in the gaps between the target layers and a nonlinear thermomechanical finite element model based on the LS-DYNA code [19] was employed to estimate the irradiation temperatures, deformations, and stresses in the capsules and contained target materials. For the encapsulated materials in either argon gas or a vacuum, heat transfer from the irradiated material contained within the capsule takes place via contact between the inner faces of the capsule windows and the target material. Contact is ensured given that there exists a two-atmosphere differential pressure for the vacuum capsule and one atmosphere for the argon-filled capsule, since irradiation is taking place in a 10-m column of water. The model capturing the conductance between the capsule window and the material layer was successfully verified against irradiated graphite and C/C composites due to the fact that these structures reveal the actual irradiation temperature during their postirradiation thermal annealing.

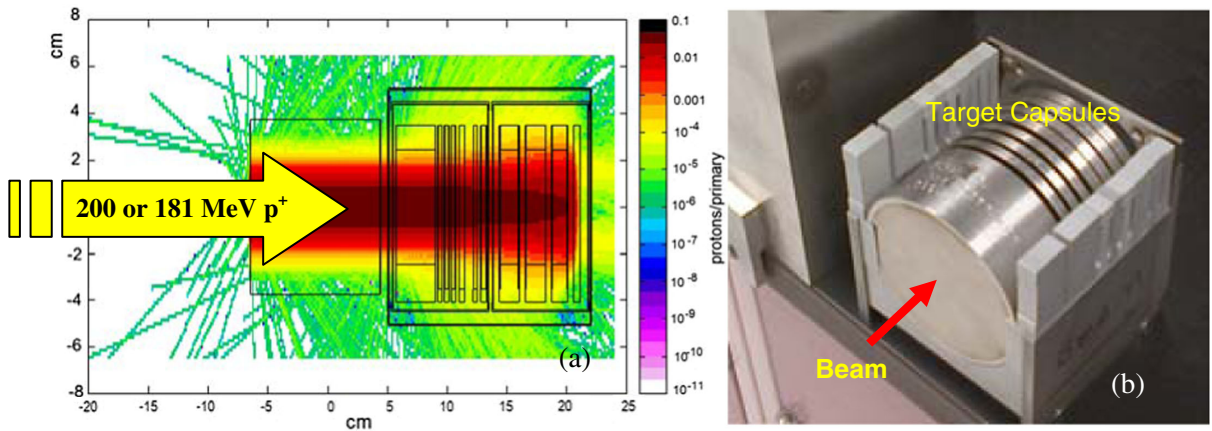

FIG. 5. Modeling of the BLIP irradiation experiment (a) and actual array setup (b).

Several irradiation campaigns of different C/C composite structures and compounds have been conducted under various environments at BLIP that included direct cooling by water and immersion in argon gas and vacuum.

Specifically, the first irradiation campaign used 200-MeV protons and evaluated low-Z pion production targets (including 3D C/C composite) for the Neutrino Factory-Muon Collider initiative [Fig. 7(a)]. The total flux of protons incident onto the 3D C/C samples which were cooled directly by the circulating BLIP deionized cooling water was $3.7 \times 10^{20}$ and the energy 130–140 MeV. Based on the Gaussian beam profile measured to have $\sigma_x = \sigma_y = 7.2$ mm, the peak fluence achieved during irradiation was $\sim 1.13 \times 10^{20}$ p/cm$^2$.

Two irradiation damage campaigns focusing on the 2D C/C structure (Toyo-Tanso AC-150) for use as a primary collimation material in the Large Hadron Collider (LHC) were conducted using the 200-MeV protons at BLIP. During the first of the two LHC 2D C/C studies [Fig. 7(b)], the total flux of protons incident onto the 2D C/C samples was $\sim 2.43 \times 10^{21}$ with the proton energy in the interval of 195–120 MeV and with the proton beam Gaussian transverse profile characterized by $\sigma_x = \sigma_y = 6.1$ mm. The peak fluence achieved during the first 2D C/C irradiation was $\sim 5.2 \times 10^{20}$ p/cm$^2$. The irradiation was conducted with 2D C/C layers cooled directly by the BLIP deionized water. The peak irradiation temperature was $\sim 120^\circ$C.

Prompted by serious structural degradation observed at the threshold fluence of $\sim 5 \times 10^{20}$ p/cm$^2$, the 2D C/C irradiation experiment was repeated with the same conditions but included for comparative purposes 3D C/C composite structures and IG43 graphite. Similarly, the irradiated array was cooled by direct contact with the deionized water and achieved a total flux of protons of $1.1475 \times 10^{21}$ under a tighter beam spot ($\sigma_x = 4.1$ and $\sigma_y = 4.5$ mm) resulting in a similar peak proton fluence as the first 2D C/C irradiation. The vulnerability of all these carbon-based structures (2D C/C, 3D C/C, and IG43 graphite) at the fluence level of $\sim 5 \times 10^{20}$ p/cm$^2$, when irradiated in water where radiolytic oxidation is aiding the structural degradation of these porous structures, was confirmed by this irradiation experiment. The accelerated degradation is attributed by the authors to radiolytic oxidation that takes place as a result of the beam-induced radiolysis of water resulting in free oxygen molecules that in turn react with carbon atoms.

An irradiation damage experiment for 3D C/C [Fig. 7(d)] that included in the test matrix several grades of graphite and hexagonal boron nitride was conducted at BNL BLIP for the Long Baseline Neutrino Facility (LBNF) of the Deep Underground Neutrino Experiment. It aimed to assess radiolytic effects on previously observed damage acceleration in carbon-based materials under energetic proton beams. To that end, irradiation with 181-MeV protons to fluence $\geq 5 \times 10^{20}$ p/cm$^2$ was conducted with the 3D C/C composite (a) in an inert gas atmosphere (argon) and (b) in direct contact with deionized cooling water. The total flux of protons incident on the LBNF array of 3D C/C and graphite was ~$3.247 \times 10^{21}$ (integrated beam current ~144 310 $\mu$A h) and the estimated peak fluence ~$6 \times 10^{20}$ p/cm². Irradiation temperatures in the array were numerically estimated (and verified via postirradiation annealing) to be ~80–100°C for the water-cooled 3D C/C and ~170–190°C for the encapsulated samples in argon gas.

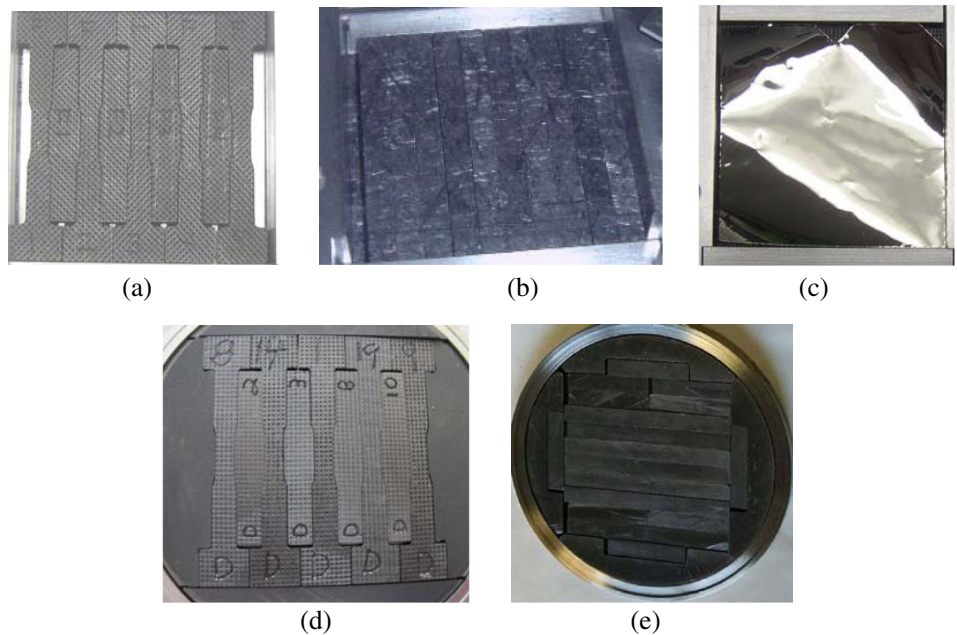

FIG. 7. Layout of C/C composite specimens for proton beam irradiation. (a) 3D C/C with fibers at 45°, (b) 2D C/C (AC-150), (c) Ni foil placed in front of a 3D C/C array for postirradiation analysis of the proton beam position and shape, (d) 3D C/C with fibers in normal orientation in argon, and (e) MoGRCF in a vacuum.

MoGRCF considered in the LHC High Luminosity beam collimation upgrade was irradiated under a vacuum using 200-MeV protons from the BNL linac to a total of ~134 000 $\mu$A h ($3.015 \times 10^{21}$ total proton flux and fluence of ~$1.1 \times 10^{20}$ p/cm²). The samples were encapsulated within special capsules [see Figs. 5(b) and 7(e)]. Peak irradiation temperatures for the MoGRCF were also numerically estimated to be ~418°C. In addition to the irradiation above, a special irradiation experiment on MoGRCF was conducted consisting of $2.8 \times 10^{18}$ p/cm² fluence at ~418°C followed by spallation neutron irradiation to a fluence of ~$3.2 \times 10^{18}$ n/cm². The combined irradiation hereafter will be referred in this paper as $6.0 \times 10^{18}$ (p + n)/cm². The spallation neutron irradiation was achieved during the isotope production with the MoGRCF capsule placed downstream of the isotope array. The 118-MeV proton beam from the linac was fully stopped within the isotope array upstream, creating a mixed irradiating field dominated by fast neutrons.

Shown in Fig. 7 are various layouts of specially fabricated C/C samples (2D and 3D C/C and MoGRCF) utilized in the irradiation campaigns. Various sample-marking schemes were used to relate each sample with a corresponding fluence and irradiation temperature, both quantities varying over the plane of the samples due to the Gaussian profile of the beam. Radiographic analysis of embedded nickel foils [Fig. 7(c)], augmented by activation measurements for each of the samples in the array, helped to establish the proton fluence in each irradiated sample. The elaborate numerical model consisting of CFD and thermomechanical analysis that was developed provide the only means for estimating irradiation temperatures.

## III. RESULTS

The postirradiation evaluation consisted of (a) a visual examination for structural degradation as a function of the fluence, (b) activation measurements and gamma spectra, (c) thermal analysis, annealing, and dimensional stability, including thermal expansion coefficient (CTE) changes as a function of the fluence, using a horizontal, two-rod high-resolution dilatometer with nanometer-level sensitivity, (d) mechanical testing (stress-strain under tension failure and load deflection), (e) ultrasonic tests for structural integrity assessment, and (e) x-ray diffraction analysis using high-energy x rays at the BNL NSLS synchrotron.

## A. Postirradiation visual assessment

A visual examination of the 2D C/C composite following the first of 200-MeV proton irradiations in deionized water with fluence $\geq 5 \times 10^{20}$ p/cm² revealed that samples receiving this threshold fluence suffered serious structural degradation. Samples receiving a lower fluence showed no signs of degradation. In order to quantify this threshold limit further and to correlate it with the response of other carbon-based structures (i.e., 3D C/C composite and IG43 graphite), the experiment was repeated under similar conditions of irradiation in water and to peak fluence above the identified threshold. Apostirradiation visual examination revealed that both composite types, 2D- and 3D-C/C, suffered degradation when exposed to a proton fluence $\geq 5 \times 10^{20}$ p/cm². Similar findings were made regarding the IG43 graphite that was irradiated along with the two composites, leading to the assessment that the observed behavior has its origin in the microstructure and that radiolytic oxidation stemming from the presence of oxygen molecules from beam-induced radiolysis coupled with their inherent porosity and the microcracking that results from the anisotropic deformation in these carbon-based structures during irradiation accelerates the structural degradation. Irradiation experiments of 3D C/C in inert gas (argon) that were subsequently performed using 181-MeV protons and with a peak fluence in excess of the identified threshold ($\geq 5 \times 10^{20}$ p/cm²) to eliminate the role of radiolytic oxidation showed that, while the 3D C/C samples maintained structural integrity overall, they exhibited signs of change revealed by the ease of material flaking off the sample.

Following the completion of the two irradiation experiments of a MoGRCF compound under a vacuum [one at $6.0 \times 10^{18}$ (p + n)/cm² and one at $1.1 \times 10^{21}$ p/cm²], a visual inspection revealed that the compound with the particular composition tested showed signs of irradiation-induced degradation even at the low fluence, while at the high fluence only samples exposed to a fluence $< 2 \times 10^{20}$ p/cm² survived. The estimated irradiation temperature of the MoGRCF during proton irradiation was ~428°C and during neutron exposure ~110°C. Figure 8 depicts images of the MoGRCF within the irradiation capsules following the tests. Images were taken through the 8-cm-thick Pb glass of the hot cell window, and so they lack clarity. The visual examination of the irradiated carbon-based structures under different environments indicated the significance of a fluence limit above which changes accelerate within the microstructure. This can be verified only with microscopic analysis. X-ray diffraction experiments conducted and discussed later in this paper attempt to link the observed macroscopic behavior with irradiation-induced changes in the lattice structure.

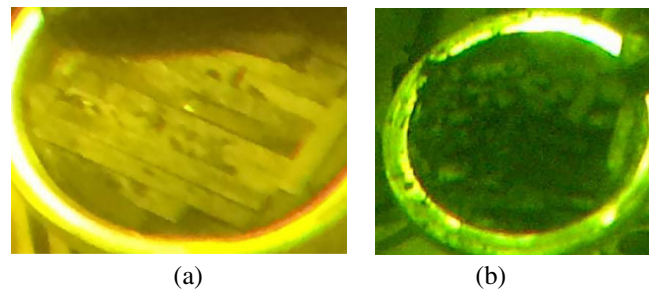

(a)                                    (b)

FIG. 8.   Observed structural degradation in irradiated MoGRCF. (a) MoGRCF irradiated with $2.8 \times 10^{18}$ p/cm² at 160 MeV followed by a fast neutron fluence of $3.2 \times 10^{18}$ n/cm². (b) MoGRCF following irradiation at a peak fluence ~$1.1 \times 10^{21}$ p/cm² with 160-MeV protons.

### B. Photon spectra

Using a high-sensitivity ORTEC Ge detector, the isotope signatures and photon spectra of the irradiated 3D C/C in argon and in contact with water were obtained and shown in Fig. 9. As can be observed, no significant difference in the photon spectrum can be identified, indicating that no radiochemistry is taking place impacting the C/C structures. Noteworthy is the $Be^7$ 477-keV peak with the same intensity for both argon- and water-irradiated samples.

### C. Stress-strain during postirradiation

The role of proton irradiation on C/C composites in changing the microstructure and leading to the increase in the ultimate tensile strength and Young's modulus has been studied through mechanical testing (tension test and three- or four-point-bending flexural tests). It should be pointed out that the three- and four-point-bending tests relate load to deflection, with the four-point bending including a larger portion of the test specimen in a pure bending state of stress. Figure 10(a) compares the unirradiated 3D C/C structure with graphite in tension showing both the repeatability of the stress-strain test and higher strength achieved with the fibers. The results of Fig. 10(b) were generated following irradiation to $1.13 \times 10^{20}$ p/cm$^2$ and depict the change in the Young's modulus with a fluence up to $0.8 \times 10^{20}$ p/cm$^2$. The tensile stress data of the irradiated 3D C/C shown are not absolute and, as noted in the Fig. 10 caption, should be used only on a relative basis to qualitatively assess changes with increased fluence. For accelerator targets or collimators exposed to beam pulses, an increase in the Young's modulus will result in an increased thermal stress in the material.

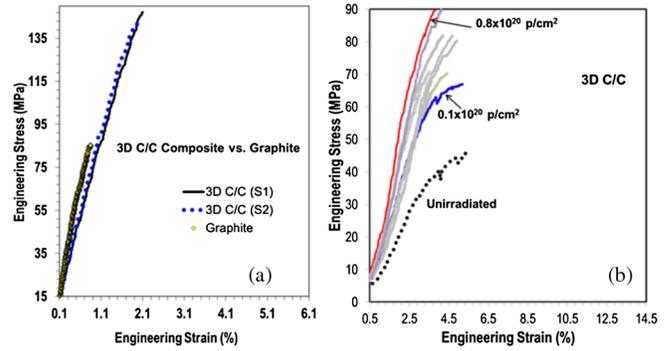

FIG. 10. Tensile tests on "dog-bone"-shaped 3D C/C proton-irradiated samples at ~100°C while directly cooled by ionized water. Shown in (a) is a direct comparison with unirradiated graphite. Slippage and premature failure at the sample head prevented the completion of the stress-strain curve (b). Test results, however, provided the basis for a radiation-induced change in the Young's modulus as a function of the fluence.

Load-deflection results of irradiated 3D C/C to a fluence of $6 \times 10^{21}$ p/cm$^2$ in terms of the load-deflection relationship are shown in Fig. 11, where unirradiated samples are compared with two samples [one of peak fluence (S1) and one of lower fluence (S2)] following irradiation in an argon atmosphere and one at peak fluence in ionized water. What is clearly observed is that the unirradiated sample (due to the relative sliding of the reinforcing fibers) experiences a large deflection before failure. The increase in the Young's modulus and ultimate tensile strength observed with an increased fluence is the due to the pinning of dislocations produced during irradiation. The radiolytic effects on the irradiated sample in ionized water are clearly visible reducing the load capacity of the cross section significantly. Figure 12 is reproduced from Nishida and Sueyoshi [4] and compares load-displacement curves for graphite and 1D and 2D C/C composites heat treated at different temperatures exhibiting similarities with the three-point-bending load-deflection results of irradiated 3D C/C composites of Fig. 11. Tension test results of irradiated 3D C/C samples produced at 45° with the horizontal (basal) fiber mesh is presented in Fig. 13. With these sample orientations, the strength of the matrix, rather than that of the fibers, was sought. The results indicate that the orientation is extremely weak (only a fraction of the strength in 0° orientation), and there is significant effect from irradiation.

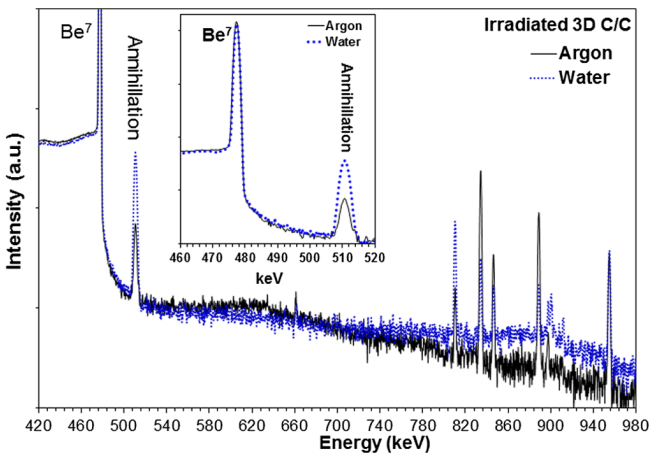

FIG. 9. Photon spectra of the C/C composite (3D) irradiated at $6 \times 10^{21}$ p/cm$^2$ in an argon atmosphere and direct contact with deionized cooling water.

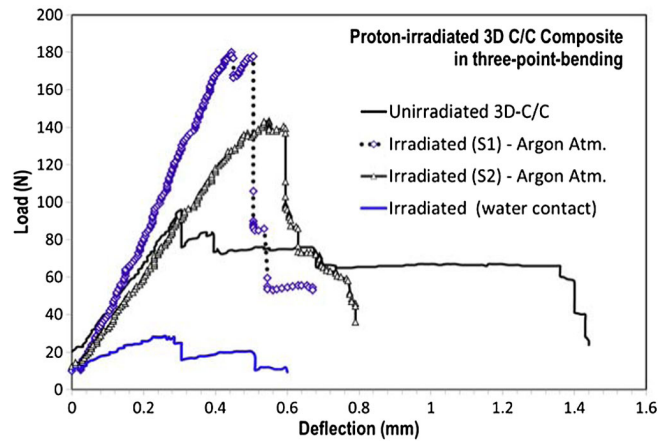

FIG. 11. Load-displacement results of 3D C/C composites irradiated in an argon atmosphere and in ionized cooling water at a fluence of ~$6 \times 10^{20}$ p/cm$^2$.

Figure 14 compares load-deflection results of unirradiated and irradiated MoGRCF compound. While the unirradiated MoGRCF response resembles that of the C/C composite, a significant change occurs following irradiation. The absence of a "continuous" net of carbon fibers in the compound leads to the cross section of the bar-type specimen (4 mm × 4 mm) exhibiting completely brittle behavior or failure. To further understand the structure of the compound, ultrasonic velocities were measured using a longitudinal transducer using the pulse echo technique and direct contact with the sample with the help of a coupler layer. The ultrasonic tests had as the objective the qualitative evolution of the microstructure resulting from irradiation. No assessment was possible for either 2D or 3D C/C structures due to the multiple reflections resulting from the fiber layers. The study, however, on irradiated MoGRCF revealed that a significant anisotropy exists in the structure of the formed compound prior to irradiation. Measurements of ultrasonic velocities showed that the velocity in the material is different in the two orthogonal directions, indicating a "weaker" and "stiffer" orientation (1530 m/s in the weak direction and 2956.5 m/s in the stiffer direction) in the structure stemming from the carbon-fiber orientation in the compound structure. Relatively low-dose irradiation of MoGRCF by 160-MeV protons (of the order of $\sim 10^{18}$ protons/cm$^2$) over a section of the specimen followed by an almost uniform mixed spectrum dominated by fast neutrons resulted in an observed increase of the ultrasonic velocity. Along the weak direction ultrasonic velocities approached 2000 m/s following irradiation or an increase of $\sim 30\%$, and along the stiffer direction ultrasonic velocities approached 3600 m/s or an increase of $\sim 21\%$.

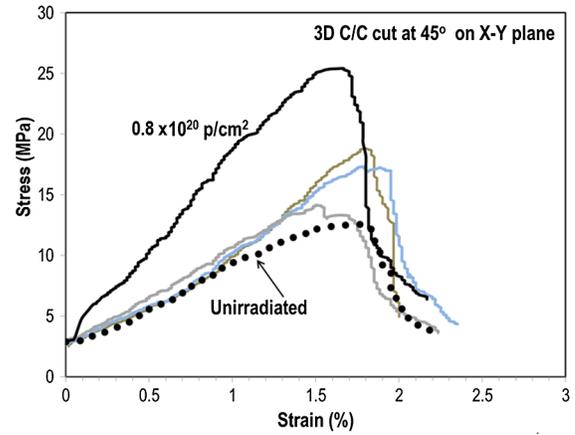

FIG. 13. Tension test of 3D C/C composite samples cut at 45° with a fiber "mesh" in the horizontal planes. The composite has <10% of the tensile strength in tension along the fibers, because it is controlled by the carbon-based fill between fibers. The effect of proton irradiation in increasing the Young's modulus and strength is more pronounced than the one observed along the fibers (Fig. 10).

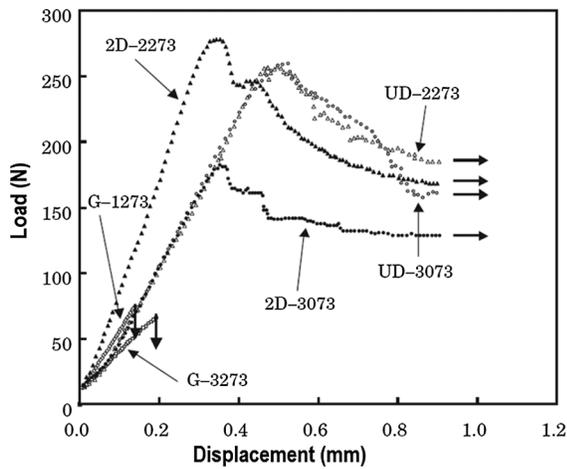

FIG. 12. Load and displacement for C/C composites and graphite. G-1273 and G-3273 indicate graphite heat-treated at 1273 and 3273 K, respectively. UD-2273 and UD-3073 represent C/C composites heat-treated at 2273 and 3073 K, respectively, whose fiber orientations are unidirectional. 2D-2273 and 2D-3073 indicate C/C composites heat-treated at 2273 and 3073 K, respectively, with 2D fiber orientations (after Nishida and Sueyoshi [4]).

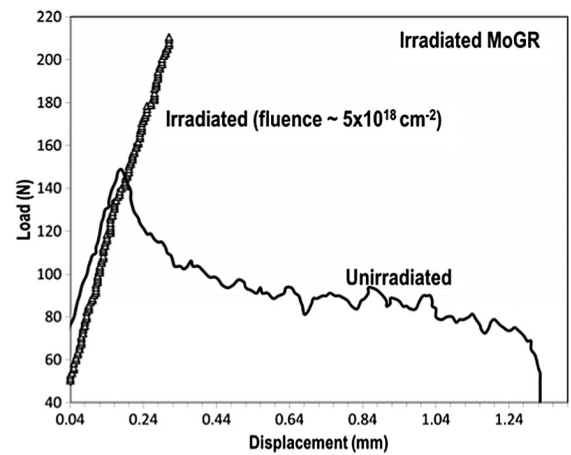

FIG. 14. Four-point-bending test comparison of irradiated and unirradiated MoGRCF. Shown is the characteristic kinking appearing to shift upwards with irradiation. Also clearly depicted is the increase in the effective Young's modulus (change of slope) following modest irradiation.

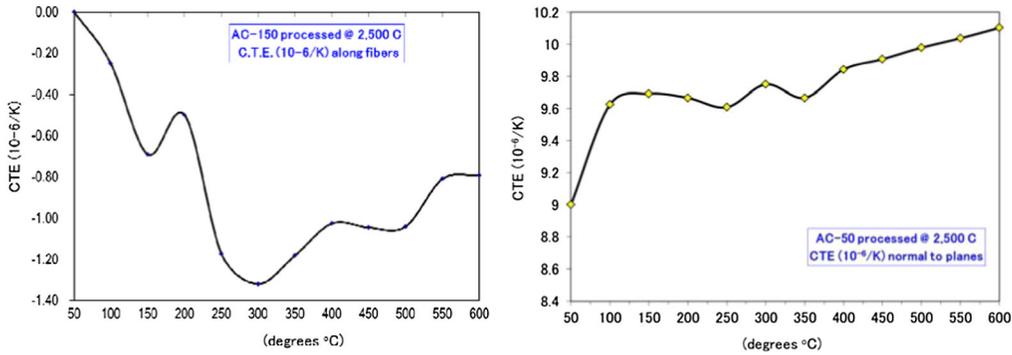

FIG. 15.  Thermal expansion coefficient of an unirradiated 2D C/C structure (Toyo-Tanso AC-150) along the two fiber orientations as a function of the temperature.

### D. Thermal stability and CTE

Thermal dimensional stability in target and collimator materials is critical to component lifetime predictions and must be understood for the informed design of beam-intercepting components. Carbon-fiber-reinforced structures are more prone to experience dimensional change because of the unique structure (fiber mesh integrated with graphitized matrix) and the high degree of anisotropy. Figure 15 depicts the CTE of an unirradiated 2D C/C structure as a function of the temperature. Observed is the significant variation with the temperature and the anisotropy that exists. A low-value, negative coefficient exists along the fibers, while normal to the planes where the structure resembles a stacking of graphite basal planes the CTE is increasingly positive. The volumetric change of the 2D C/C structure with the thermal cycling was studied with the results shown in Fig. 16. It is observed that the structure stabilizes following thermal annealing in both a longitudinal and normal to the planes orientation. This results from the reduction of manufacturer porosity by thermally induced growth into the preexisting pores.

The results of the postirradiation annealing of the AC-150 (2D C/C) irradiated with 200-MeV protons to $4.5 \times 10^{20}$ p/cm$^2$ are shown in Fig. 17. What is observed during postirradiation annealing is that normal to the fiber planes the structure contracts during the first thermal cycle at temperatures above the irradiation temperature due to the annealing of interstitials in the lattice of the graphitized matrix (filler). The opposite effect is experienced along the fiber planes, where the structure expands as a result of defect annealing. Important to note is that, during postirradiation annealing, a fraction of the irradiation-induced dimensional change is recovered. The annealing temperature above the irradiation temperature represents activation energy for interstitial atoms at a certain distance from the basal plane to return to empty sides in the octahedral planes causing a dimensional increase along the plane. When annealing at a higher temperature, no dimensional change is observed up to the previously achieved temperature due to the fact that all interstitials that could be activated and return to sites on the basal plane have done so. Subsequently, volumetric change begins again in excess of the previous postirradiation annealing temperature.

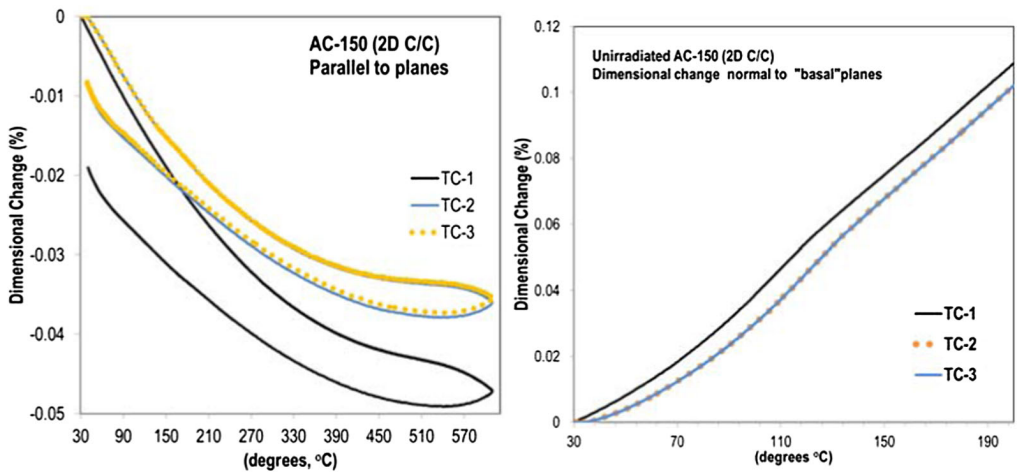

FIG. 16.  Preirradiation thermal cycle annealing of the 2D C/C structure (Toyo-Tanso AC-150) along the two fiber orientations.

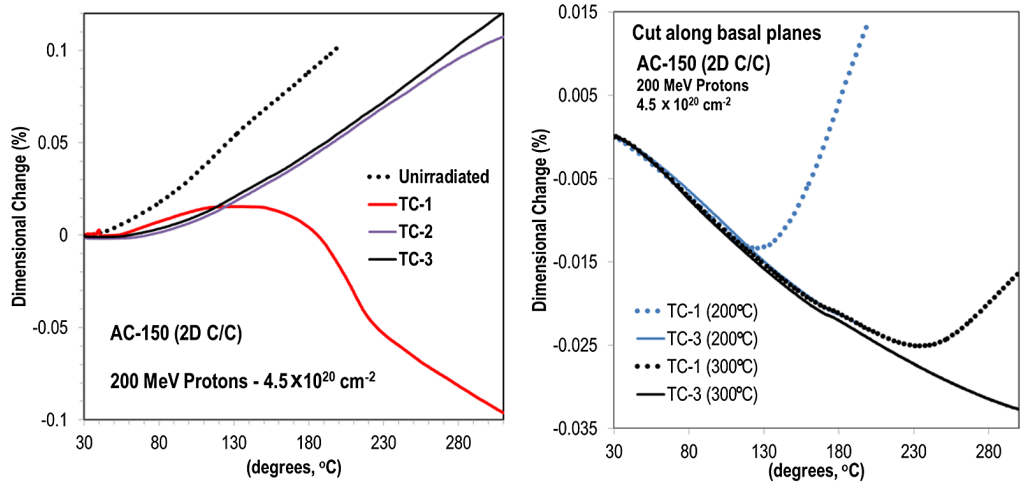

FIG. 17.   Postirradiation thermal cycle annealing of the 2D C/C structure (Toyo-Tanso AC-150) along the two fiber orientations: normal to the fiber planes (left) and along the fibers (right). TC- (200°C) and TC- (300°C) indicate thermal cycles to peak temperatures 200°C and 300°C, respectively.

The pre- and postirradiation annealing of 3D C/C has been studied following the irradiations with 200- and 181-MeV protons. Shown in Fig. 18 is the "adjustment" of the structure resulting from manufacturing porosity and defects. This prompts the recommendation that, prior to use of these composite structures as targets or collimators where thermal stability is crucial (especially for collimators), the materials should undergo thermal annealing or thermal cycling to temperatures equal to or greater than the anticipated operating temperature. The evolution of thermal expansion coefficient with postirradiation annealing and its dependence on the average proton fluence received by 3D C/C samples irradiated in the same array with a similar fluence (samples from two different layers along the proton beam path) is shown in Fig. 19. The effects of a higher dose ($6 \times 10^{20}$ p/cm²) on the dimensional stability of the 3D C/C structure in combination with the irradiating environment (argon atmosphere vs water) are shown in Figs. 20, 21, and 22. The annealing response is similar to the postirradiation response exhibited by the 2D C/C structure along the fibers; i.e., above the irradiation temperature, the fiber-filler matrix undergoes an expansion due to the mobilization of interstitials and their return to empty sites in the lattice and on the octahedral planes. The effect of the environment after the first thermal cycle is minimal. During the first postirradiation cycle, annealing in the material irradiated in contact with water begins earlier due to the fact that the irradiation temperature was lower. The inflection point of the curve reveals the irradiation temperature for the two environments. The effect of higher activation energies (i.e., higher annealing temperatures) on the progressive dimensional change of a highly irradiated 3D C/C composite in argon is shown in Fig. 22 and resembles the annealing behavior of the 2-D C/C along the fiber planes.

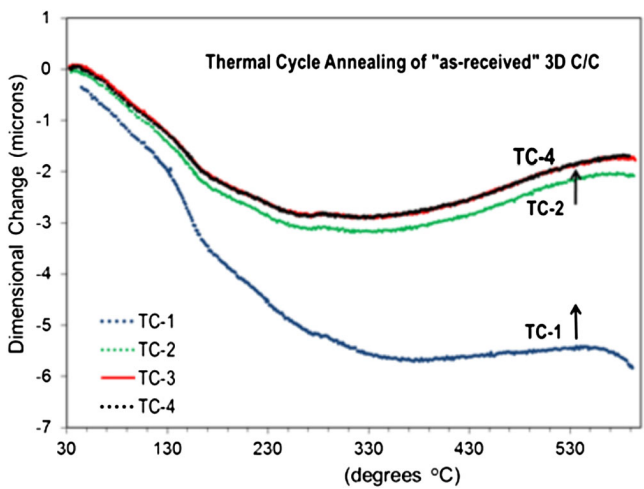

FIG. 18.   Preirradiation thermal cycle annealing of the 3D C/C structure showing achieved dimensional stability after the third thermal cycle.

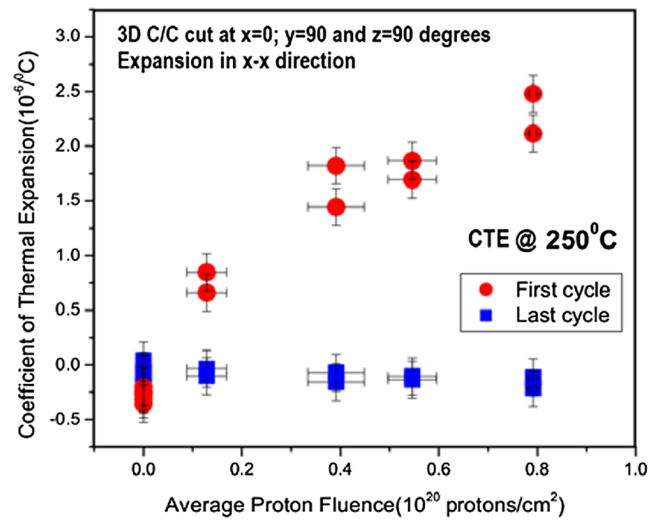

FIG. 19.   Thermal expansion coefficient of the proton-irradiated 3D C/C composite as a function of the average fluence. Two different samples irradiated at a similar fluence were tested and shown.

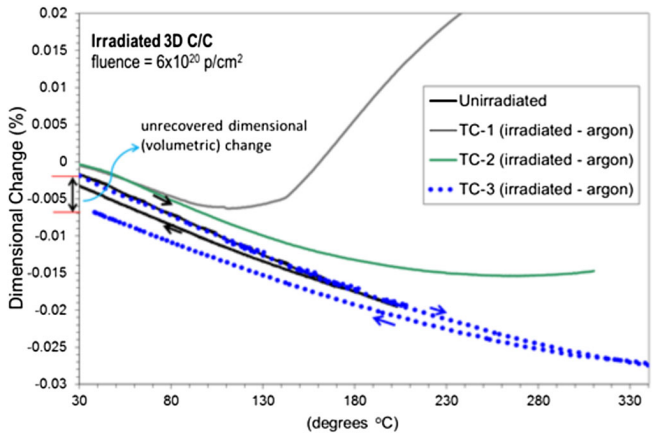

FIG. 20. Postirradiation thermal annealing of irradiated 3D C/C to a high dose in an argon atmosphere.

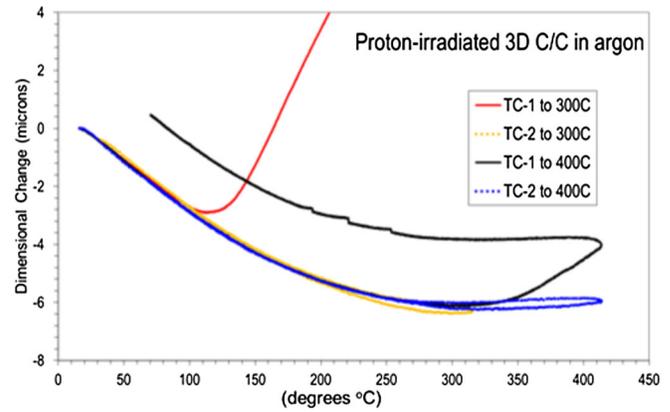

FIG. 22. Postirradiation thermal cycle annealing with an increasing cycle temperature (300°C vs 400°C) of proton-irradiated 3D C/C to a high dose.

Dimensional changes in an irradiated MoGRCF compount including postirradiation annealing are shown in Figs. 23 and 24. It is observed that the dimenensional change in as-received MoGrCF as a function of the temperature for the temperature range is lower than that of the fine-grained IG-430 graphite and 2D C/C and exhibits a distinct inflection point at ~500°C. Following proton irradiation up to $1.5 \times 10^{20}$ p/cm$^2$ (which is the fluence of samples in the vacuum capsule that remained intact during irradiation [Fig. 8(b)]), the compound MoGRCF exhibits a lower thermal expansion than pure graphite. Upon thermal cycle annealing to temperatures up to 700°C, the MoGRCF compound stabilizes at levels below those of the as-received material. A complete thermal cycle (heating and cooling), TC-3, of irradiated MoGRCF is also depicted in Fig. 24 showing a "stable" structure following the annealing of damage achieved during the TC-1 and TC-2 postirradiation thermal cycles to 700°C.

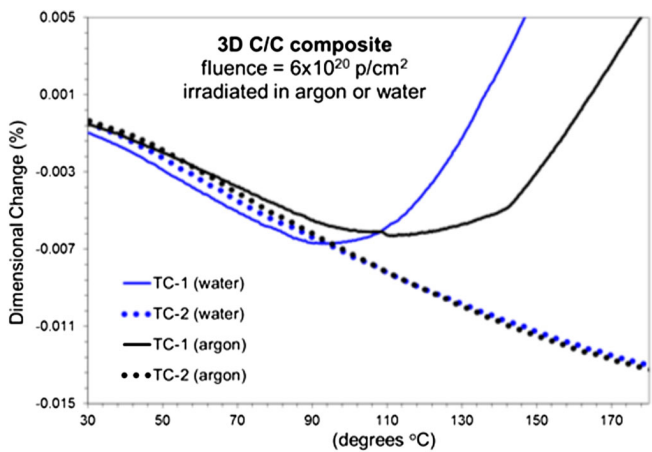

FIG. 21. Postirradiation thermal cycle annealing of proton-irradiated 3D C/C to a high dose comparing the effects of the irradiating environment (argon vs water).

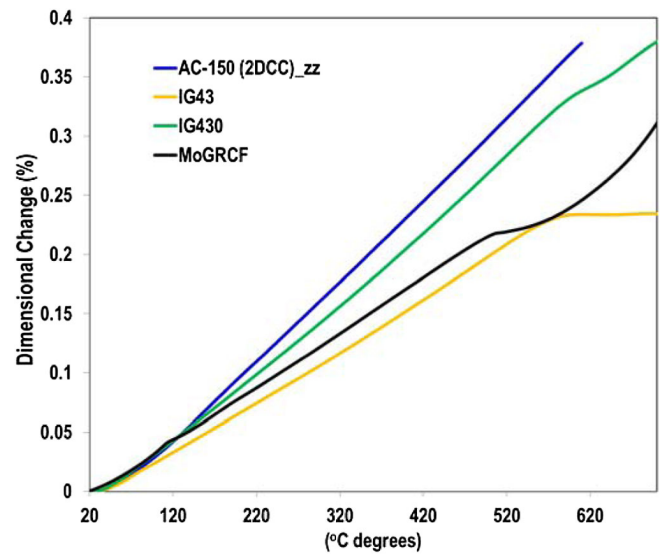

FIG. 23. Comparison of the dimensional change in unirradiated MoGRCF with IG-43, IG-430, and 2D C/C.

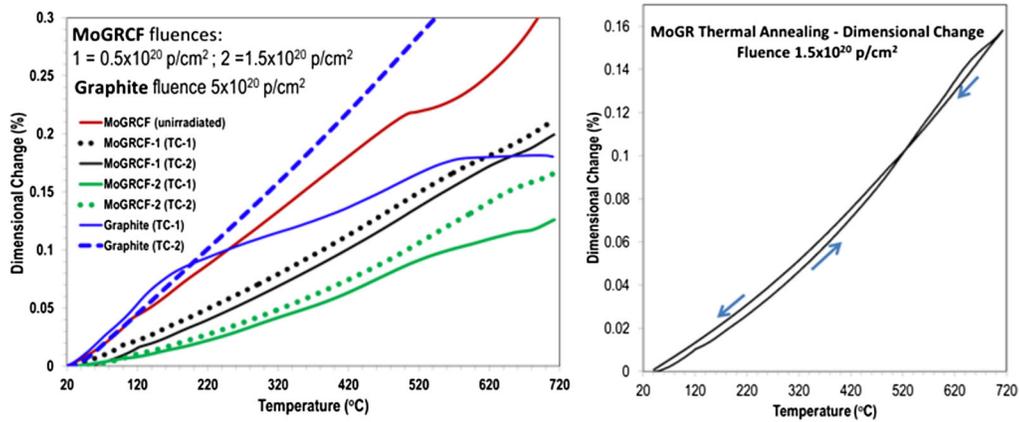

FIG. 24. Postirradiation annealing of MoGRCF compared to graphite (left) and the dimensional change experienced by irradiated MoGRCF during a complete thermal cycle (right). Noted is the crossover during the cooling phase of the cycle.

## E. X-ray diffraction studies of proton-irradiated C/C composites

To shed light on the microstructural changes occurring during the proton irradiation of carbon composite structures and the MoGRCF compound that manifest themselves macroscopically as a structural degradation or changes in physical properties, x-ray diffraction experiments were conducted using high-energy x rays at the BNL synchrotron. The energy dispersive x-ray diffraction (EDXRD) technique with 200-keV white beam x rays was performed at the X17B1 beam line and the XRD technique using 70-keV monochromatic x rays at the X17A beam line. A specially designed experimental stage facilitated an *in situ* four-point-bending state of stress during x-ray irradiation. EDXRD, an analytical technique for characterizing materials, differs from conventional x-ray diffraction in that it uses polychromatic photons and operates at a fixed angle while enabling the collection of full diffraction patterns. For these experiments, EDXRD facilitated the use of the experimental stage providing *in situ* four-point-bending stress while the sample was irradiated with x rays.

The primary objectives of the two-phase x-ray diffraction study were to (i) obtain insight as to why there appears to be an acceleration of structural integrity degradation above the fluence of $5 \times 10^{20}$ p/cm$^2$ at these proton energies (120–200 MeV) of carbon-based structures and are more pronounced when radiolytic oxidation is involved—specifically, experimentally assess whether such behavior is directly connected with the lattice structure changes at this particular fluence threshold and possibly correlate it with experimental evidence at these fluence levels from other irradiating species (i.e., experience with neutrons); (ii) explore whether the various types of C/C composite structures show such a similar trend, are subject to similar threshold fluences, and how they perform compared to graphite—assess the potential role graphitization achieved during manufacturing; and (iii) assess why the tested MoGRCF compound grade appears to be as vulnerable to irradiation even at fluences as low as $\sim 6 \times 10^{18}$, where it already exhibited signs of structural distress and such a dramatic change during load-displacement tests. More importantly, why is it so vulnerable to total destruction above a threshold fluence (threshold for MoGRCF lower than the observed threshold for both 2D and 3D C/C composites and graphite under radiolytic oxidation). We have estimated that the limit is $\sim 1.5 \times 10^{20}$ p/cm$^2$, since the samples in the array structurally at least survived. Important to note is that no environmental effects during MoGRCF irradiation were contributing to the observed damage (irradiation in vacuum) while the irradiation temperature was estimated to be >400°C, which should have induced some degree of damage annealing. It is therefore anticipated that the damage origin is in the microstructure of the compound.

Graphitization is the process of converting amorphous carbon into graphite by heat treating the material to temperatures above 2500°C. During the process, the carbon which is in a turbostratic state (less-ordered) is converted into an ordered three-dimensional graphite structure. With graphitization, the lattice order is increased, accompanied by a decrease in intralayer distances. Irradiation damage, on the other hand, will alter the graphite lattice structure by displacing atoms from the basal planes and forming interstitials to clusters between the planes, leading to an increase of the interplanar distance and the broadening and shifting to lower diffraction angles of the characteristic peaks (002) and (004) with a simultaneously observed reduction of the peak intensity. With increased irradiation, coalescence of interstitial clusters and formation of new, less-ordered crystalline planes (turbostratic state) are formed between the basal planes. The collapse of vacancies formed within the basal planes will lead to the shrinking of the interplanar distance characterized by the shifting of the diffraction angle of (112) to higher values. The initial state of the C/C composites and MoGRCF compound (including graphite for reference or comparison) and the evolution of the lattice structure with irradiation fluence have been assessed and reported.

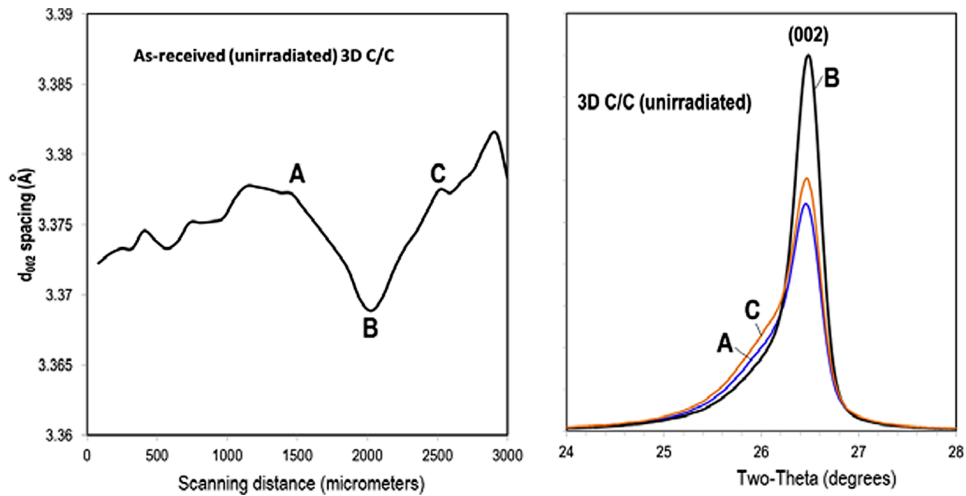

FIG. 25. D spacing and (002) diffraction peak of an unirradiated 3D C/C composite.

### 1. X-ray diffraction of 2D- and 3D-C/C composites

Selected samples of 2D and 3D C/C composites ($42 \text{ mm} \times 4 \text{ mm} \times 4 \text{ mm}$) irradiated to a high fluence ($2 \times 10^{20}$ and $6 \times 10^{20}$ p/cm$^2$, respectively) had been studied using the EDXRD technique at the X17B1 beam line of NSLS. The 2D C/C was irradiated in contact with cooling water, and it was below the threshold fluence that induced structural degradation. The 3D C/C composite had been irradiated in an argon atmosphere and macroscopically, other than the ease of material flaking off from the surface, seemed to maintain overall structural integrity. As noted previously, in the unirradiated (as-received) state, it is expected that different degrees of graphitization in the carbon fibers and the carbon matrix are achieved during the manufacturing process. How that, if at all, affects the evolution under irradiation is a subject of this study.

Furthermore, irradiation damage to the carbon-based structures (fibers and matrix) will manifest itself in the broadening of the (002) and (004) peaks indicating radiation-induced amorphization with the appearance of new crystalline planes (turbostratic state) between basal planes due to the coalescence of interstitial clusters, reduction of the diffraction peak intensities, and increase in the interplanar distance $d_{002}$ along the $c$ direction (or shifting of the diffraction angle $2\theta$ to lower values).

Shown in Fig. 25 is the variation of the d spacing ($d_{002}$) and the (002) diffraction peak of as-received, unirradiated 3D C/C composite demonstrating the variation in the degree of graphitization that exists in the structure as a result of the presence of fibers integrated with the matrix (the higher the degree, the lower the d spacing). This is also shown as a knee on the side of the diffraction peak towards lower $2\theta$.

The 3D EDXRD phase map of the 3D C/C unirradiated sample and the sample irradiated to $6 \times 10^{20}$ p/cm$^2$ is shown in Fig. 26, showing peak broadening and peak shifting as well as disordering. A more detailed comparison is depicted in Fig. 27, where the evolution of the diffraction peaks (002) and (004) is captured. Shown in Fig. 27(a) are the change in the lattice parameter $d_{002}$ ($\sim 2.81\%$) and the graphitization induced by irradiation. The latter is deduced from the disappearance of the knee on the left of the (002) diffraction peak, a finding that is significant for the C/C composite implying, at least qualitatively, that the initial degree of graphitization may not matter at all. Besides (002) peak broadening due to irradiation, one clearly observes the appearance of new crystalline planes (knee on the right side of the 002 diffraction peak) resulting from interstitial clusters coalescing (turbostratic state). A significant change in the (004) diffraction peak at the fluence of $6 \times 10^{20}$ p/cm$^2$ is evident in Fig. 27(b), indicating along Fig. 27(a) that the structure is becoming more amorphous. To compare with the effects observed in graphite under similar fluences and the same environmental conditions, the evolution of the (002) diffraction peaks (along with the $d_{002}$ lattice parameter) which relate to the damage by the formation of interstitial clusters leading to growth of the crystals in the $c$ direction (shifting to lower $2\theta$) is shown in Fig. 28(a). The irradiation-induced changes on the basal planes are associated with vacancy cluster formations which lead to a shrinking of the lattice constant in the plane and manifested as a shifting of the (112) diffraction peak to higher $2\theta$ values as seen in Fig. 28(b) for both carbon-based structures (graphite and 3D C/C).

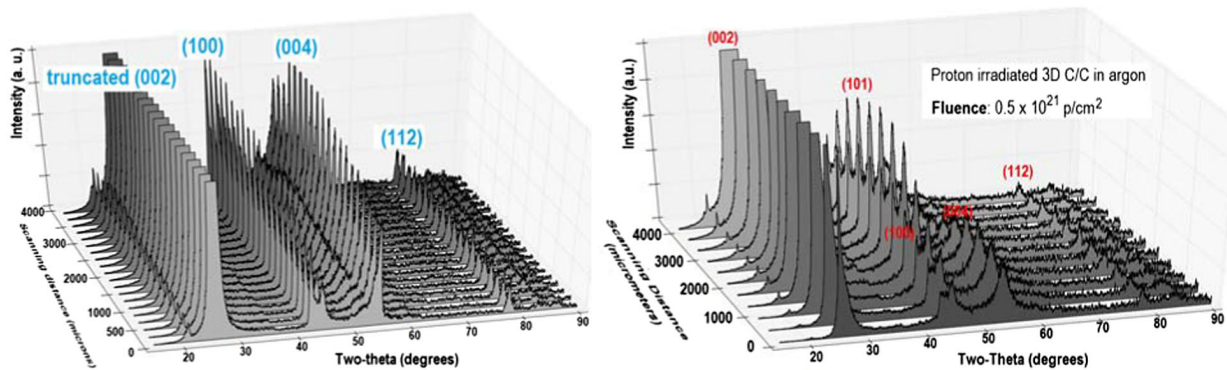

FIG. 26. 3D EDXRD phase map of as-received (left) and irradiated 3D C/C composite (right) to $6 \times 10^{20}$ p/cm$^2$ in an argon atmosphere. Peak broadening and disordering are due to defect cluster formation.

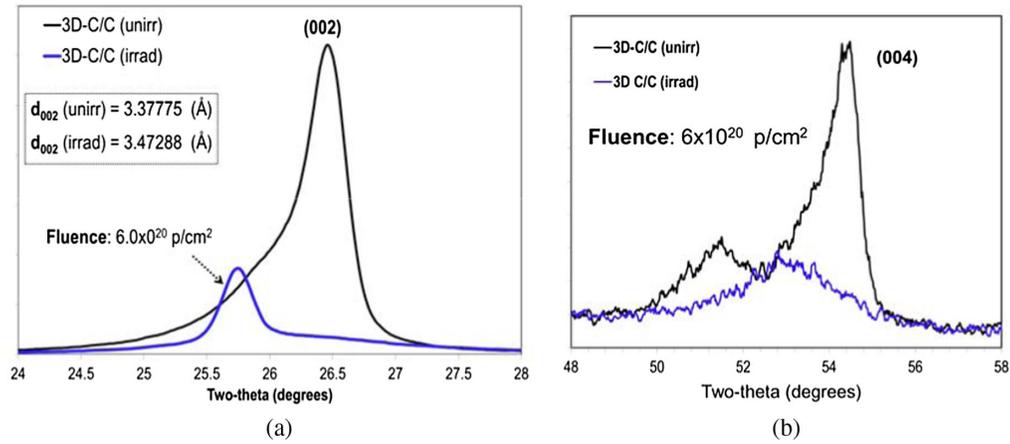

FIG. 27.   Irradiation-induced interplanar change and graphitization (incomplete graphitization indicative by the knee on the LHS of the unirradiated curve) of fibers in the 3D C/C composite.

Figure 29 depicts postirradiation x-ray diffraction results of the proton-irradiated 2D C/C composite structure to the fluence of $2 \times 10^{20}$ p/cm$^2$ and its direct comparison with the 3D C/C counterpart. The AC-150 or 2D C/C composite differs from the 3D structure in that it consists of stacked layers of weaved fibers along the plane separated by a graphitized matrix. What is important to observe in the 3D phase map of Fig. 29(a) is the appearance at the fluence of $2 \times 10^{20}$ p/cm$^2$ of new crystalline planes as a result of interstitial cluster formation indicating the start of turbostratic state. Figure 29(b) depicting the d-spacing fluctuation when scanning with x rays parallel to the fiber planes and over the specimen thickness ($d_{002}$ represents the distance between graphite planes) clearly shows the variation between the fibers and the matrix (higher d-spacing indicates a lesser degree of graphitization). Figures 29(c) and 29(d) compare the XRD spectra between 2D and 3D C/C structures for the (002), (100), (101), and (002) planes. As noted in other studies [2], lower graphitization is expected in the fibers than the graphitized matrix which in turn indicates that the interplane distance is larger in the fibers than in the matrix.

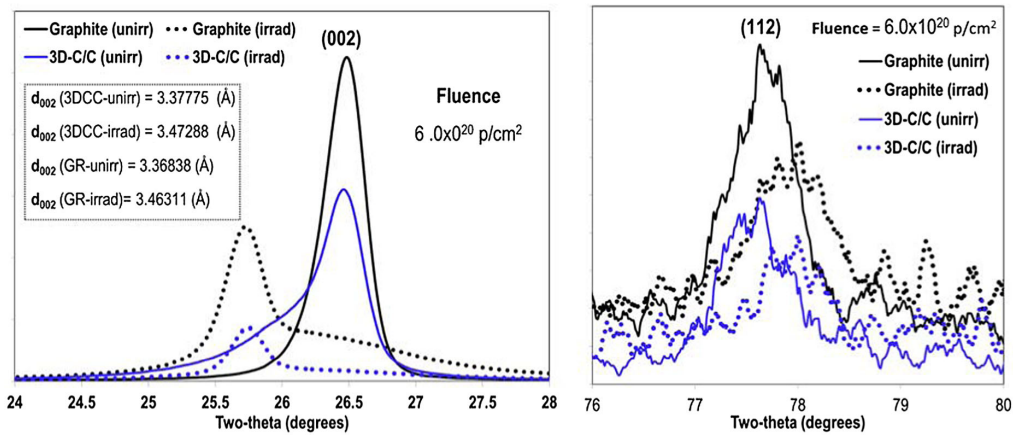

FIG. 28.   Irradiation-induced changes in the (002) and (112) diffraction peaks in 3D C/C compared to irradiated graphite with a similar proton fluence. Both graphite and 3D/C were irradiated in argon gas.

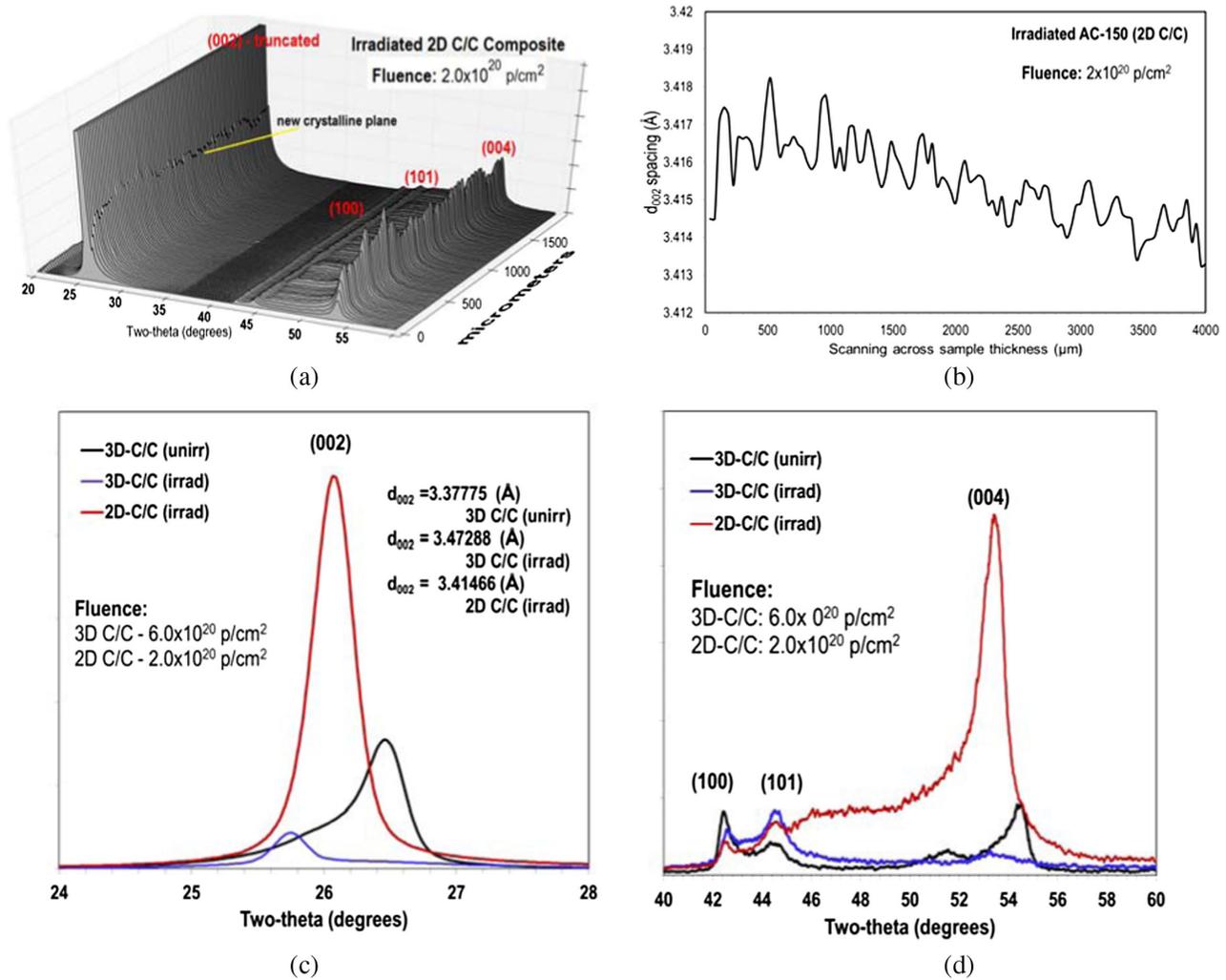

FIG. 29.   Irradiation-induced changes in the XRD spectra and d spacing of irradiated 3D C/C and 2D C/ composites irradiated to different proton fluences.

In an effort to further qualify and/or quantify the accelerated damage observed at fluences in excess of $5 \times 10^{20}$ p/cm$^2$ and, in particular, following irradiation combined with radiolytic oxidation and link it with changes in the microstructure, experience data from fast neutron irradiation on graphite are used for reference. As shown in Fig. 30(a) (reproduced from Ref. [16]), fast neutron effects on the lattice structure of graphite experience a dramatic increase shown by a jump in (002) peak broadening implying amorphization (turbostratic state). The present study confirms for the first time [as shown in Fig. 30(b)] that energetic protons induce a similar transition in graphite and 3D C/C irradiated in an inert gas atmosphere. This further explains the role of radiolytic oxidation in the observed structural degradation of 2D C/C, 3D C/C, and graphite structures upon exceedance of $5 \times 10^{20}$ p/cm$^2$. At these fluence levels the carbon-based structure becomes more amorphous and porous, thus providing increased reaction surfaces for oxygen molecules generated from the radiolysis of water due to the ionizing beam interacting with carbon atoms.

These findings, based on lattice changes as a function of the received fluence, provide the most accurate metric in establishing operational limits of these carbon-based structures under energetic protons.

### 2. X-ray diffraction of MoGRCF

Two capsules of the MoGRCF compound had been irradiated in a vacuum. The sample that received a peak

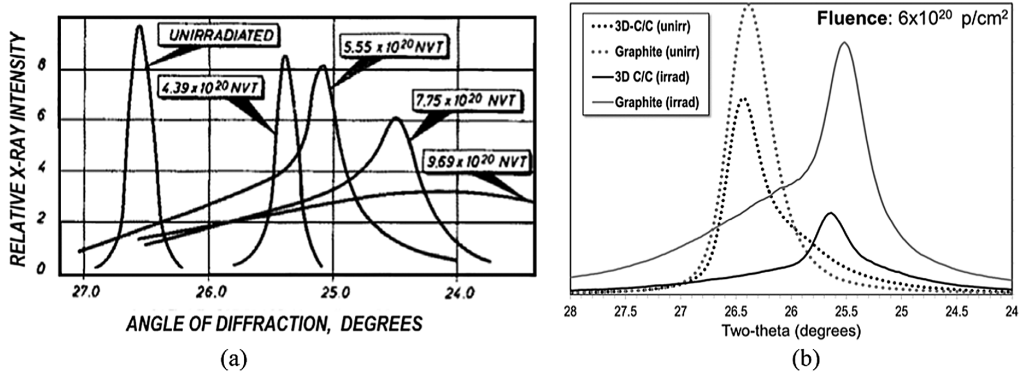

FIG. 30. Comparison of the effects of fast neutrons on graphite as a function of the fluence [16] with the effects on graphite and 3D C/C composite structure by energetic proton irradiation in inert gas (argon).

fluence $\sim 5 \times 19^{20}$ p/cm$^2$ experienced serious structural degradation above $1.5 \times 10^{20}$ p/cm$^2$ [as shown in Fig. 8(b)]. The other sample was exposed to a fluence of $\sim 6 \times 10^{18}$ (p + n)/cm$^2$ [shown in Fig. 8(a)]. This latter sample was studied only using the high-energy synchrotron x rays (monochromatic and polychromatic) due to the dose limitations imposed by the synchrotron. Specifically, the dose of surviving samples from the first capsule exposed to $\leq 1.5 \times 19^{20}$ p/cm$^2$ far exceeded the radioactivity limits of the synchrotron facility. The primary objective was to examine the evolution of the microstructure to possibly explain the distress already visible in this grade of MoGRCF [Fig. 8(a)] exposed to fluence as low as $6 \times 10^{19}$ cm$^{-2}$. In addition to the visual signs of beginning

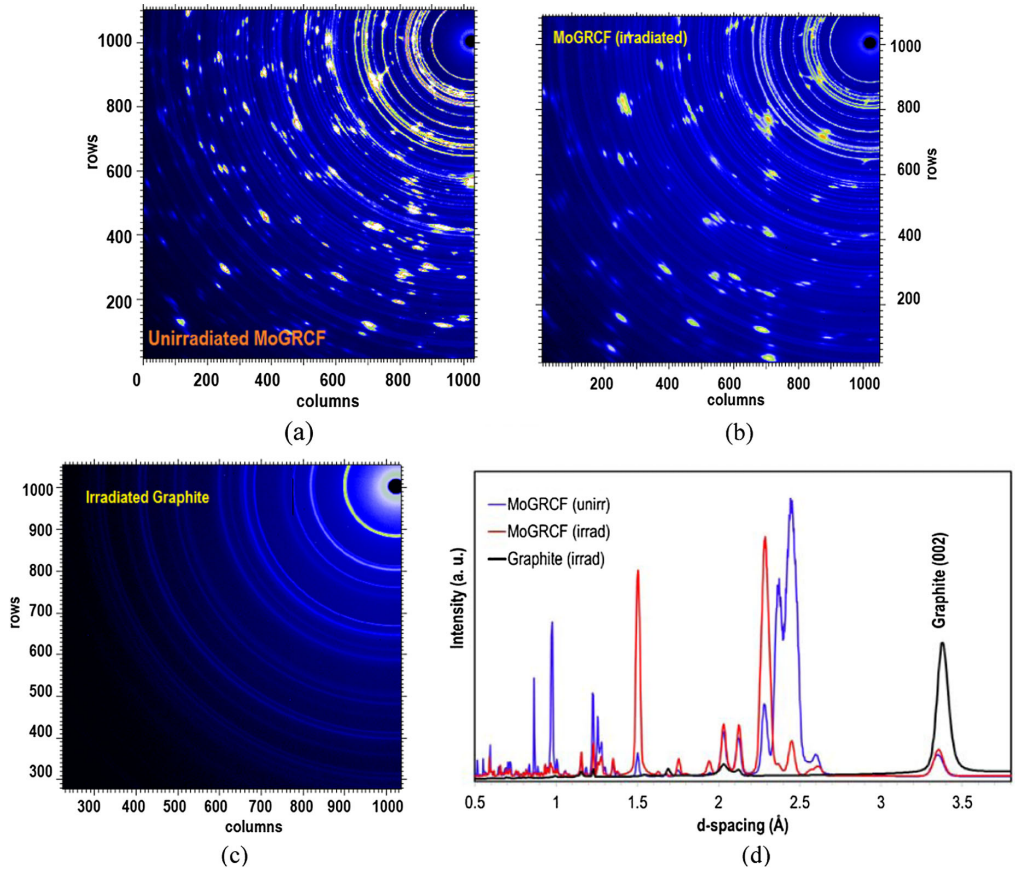

FIG. 31. Diffraction pattern of (a) unirradiated MoGRCF, (b) irradiated to a modest fluence [$\sim 6 \times 10^{18}$ (p + n)/cm$^2$], and (c) irradiated graphite ($2 \times 10^{19}$ n/cm$^2$), and (d) d spacing extracted from the diffraction patterns of the 2D area detector images [(a)–(c)]. One-quarter of the detector is shown in (a)–(c).

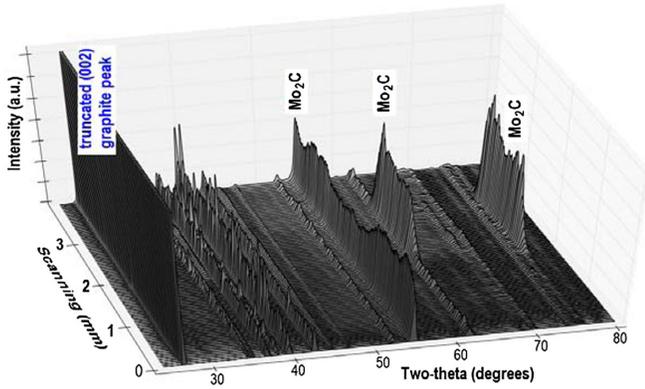

FIG. 32.　3D phase map of unirradiated MoGRCF generated using 200-keV polychromatic x rays via the EDXRD technique showing an uneven distribution of $Mo_2C$ phases from manufacturing of the compound.

degradation, the vulnerability of the compound was further confirmed during postirradiation mechanical tests where a completely brittle behavior was observed. Two x-ray diffraction experiments were conducted on MoGRCF (for both unirradiated and irradiated samples). Specifically, monochromatic 70-keV and polychromatic (white) 200-keV x rays were used to assess both the preirradiated (as received) and the radiation-induced microstructural evolution of the MoGRCF compound. The objective was to assess whether the state of the compound prior to irradiation, specifically the distribution of phases present in the compound such as graphite, molybdenum, and molybdenum-carbide(s), may explain the vulnerability exhibited. Shown in Fig. 31 are 2D diffraction patterns of unirradiated and irradiated MoGRCF as well as of irradiated graphite (for comparison purposes) obtained in the transmission mode using the 70-keV monochromatic x rays at the X17A beam line of NSLS. The d spacing of the three materials extracted from the diffraction patterns of Figs. 31(a)–31(c) revealing the phases and their evolution is depicted in Fig. 31(d). The x-ray patterns depicted in

Fig. 31 reveal some very important facts regarding the grade of MoGRCF tested. Specifically, the unirradiated compound shows grain inhomogeneity with spots revealing larger grains of intermediate phases. Following a modest irradiation with a fluence $\sim 6 \times 10^{18}$ $(p+n)/cm^2$, an irradiation-induced transformation of intermediate phases towards the $Mo_2C$ phase accompanied by the reduction of crystallinity is observed. In addition, it is clear from the graphite (002) diffraction peak that the graphite phase is well developed during the fabrication of the compound, thus excluding the possibility that the vulnerability stems from incomplete graphitization of the phase.

The 3D phase map of as-received MoGRCF obtained using the 200-keV polychromatic x-ray beam of the X17B1 beam line is shown in Fig. 32, clearly demonstrating the inhomogeneity of phases in the manufactured compound and further validating the data of Fig. 31(a) obtained with monochromatic x rays. Figure 33 depicts XRD spectra of unirradiated MoGRCF compared with graphite and 3D C/C. The XRD spectra shown in Fig. 33 confirm that the premature failure of the studied MoGRCF grade was not the result of poor graphitization achieved during the manufacturing of the compound, since its graphitization surpasses the other two structures (narrower 002 diffraction peak and lowest $d_{002}$). Specifically, measured $d_{002}$ values of the three unirradiated material structures shown in Fig. 33 are 3.35123 (Å) for MoGRCF, 3.36838 (Å) for graphite, and 3.37775 (Å) for 3D C/C composite.

Shown in Figs. 34(a)–34(c) are the effects that a modest irradiation $[6.0 \times 10^{18}$ $(p+n)/cm^2]$ has on the microstructure of MoGRCF. The irradiated MoGRCF sample examined was exposed to $\sim 3.2 \times 10^{18}$ $n/cm^2$ uniformly throughout and to an additional $\sim 2.8 \times 10^{18}$ $p/cm^2$ Gaussian beam at the center, thus offering a variation of the fluence across the sample. Shown in Fig. 33(a) are diffraction data while scanning across the zone of peak fluence. The appearance or enhancement and disappearance of Mo-C phases shown over the zone exposed to a higher dose is evident and further confirms the findings depicted in Fig. 31(b) of a radiation-induced transformation of the intermediate phases. The effect of these modest irradiation levels on Mo-C and the graphite phases of the MoGRCF compound are shown in Fig. 34 based on a scan at the center portion of the 40-mm-long sample exposed to the peak fluence of $6.0 \times 10^{18}$ $(p + n)/cm^2$.

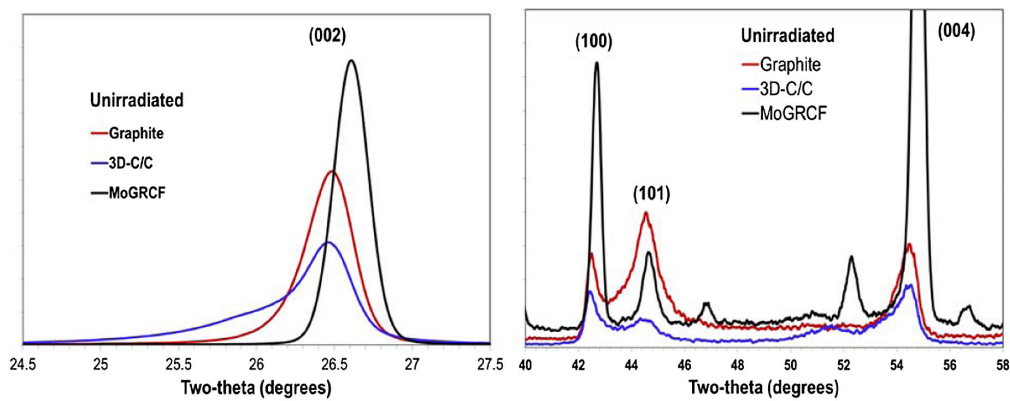

FIG. 33.　Comparison of XRD spectra of unirradiated MoGRCF, 3D C/C, and graphite.

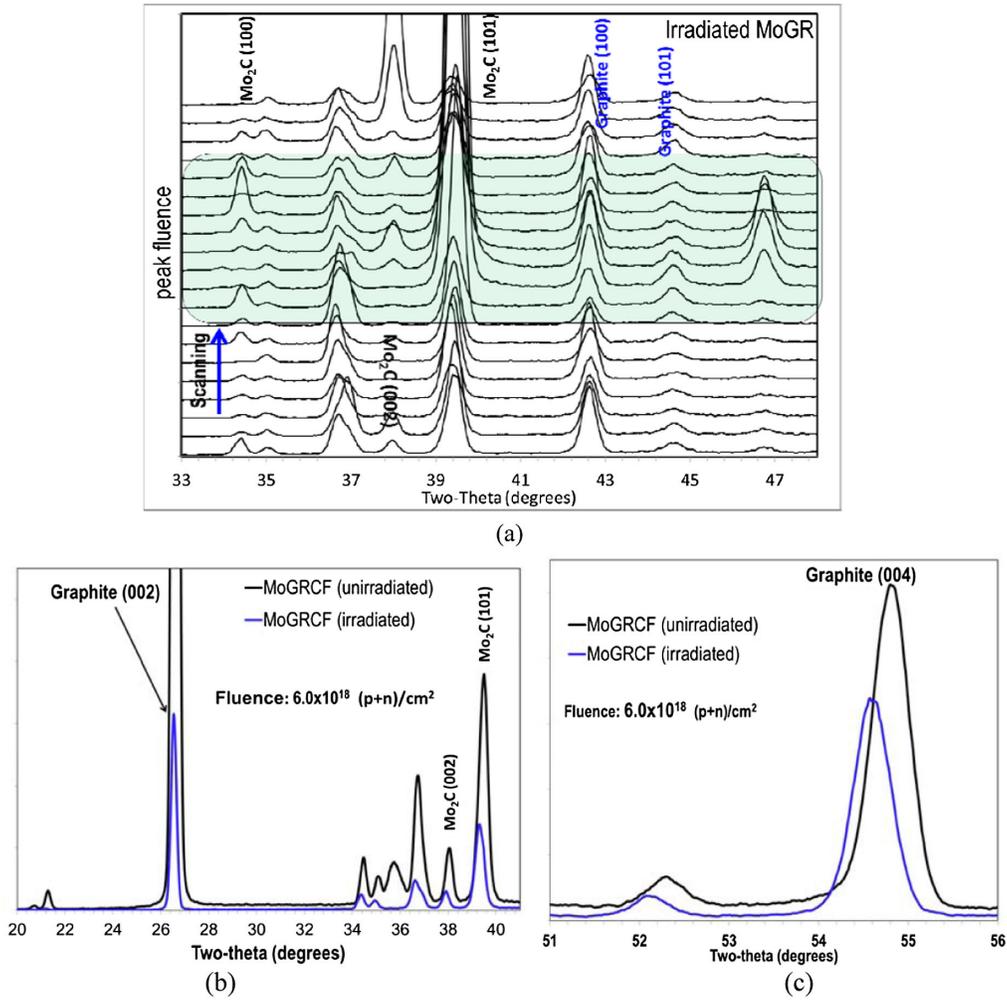

FIG. 34.  Irradiation-induced changes in MoGRCF observed via the EDXRD technique and 200-keV white x-ray beam: (a) scanning through a region of peak fluence $6 \times 10^{18}$ $(p + n)/cm^2$, (b) evolution of phases in the compound, and (c) radiation-induced changes in the graphite phase (004) diffraction peak indicating crystal growth.

## IV. SUMMARY, DISCUSSION, AND FUTURE WORK

A multifaceted study consisting of high-energy beam shock studies and a series of irradiation damage experiments with proton beam energies in the range of 120–200 MeV of C/C composites and a MoGRCF compound were conducted. The primary objectives were to assess how C/C composites respond to intense beam pulses as compared to graphite, how resilient during prolonged irradiation, and, in particular, how dimensionally stable, they are, and whether fluence limits can be identified. For the new compound MoGRCF irradiated at a high fluence, for the first time we assess how stable it is under irradiation and evaluate its potential as a collimator material in the high luminosity LHC. Macroscopic postirradiation analyses of thermal stability, CTE evaluation, mechanical behavior, and annealing were conducted at the BNL Hot Cell Facility. Microscopic postirradiation evaluations using x-ray diffraction techniques were conducted at the BNL NSLS synchrotron using high-energy polychromatic and monochromatic x-ray beams. The primary objective of the x-ray diffraction studies was to qualify the macroscopically observed behavior of these materials above certain fluence thresholds and was part of the effort to establish operational limits. Off-beam studies assessing the microstructural response of C/C composites and the MoGRCF metal matrix compound to high temperatures were also undertaken.

The study has revealed the following: (i) 24-GeV shock experiments confirmed the superior, compared to that of graphite, ability of the 3D C/C fiber-reinforced structures to absorb a beam-induced shock, while the experimental data deduced provide an excellent basis for the calibration of predictive numerical models for an intense beam response. (ii) As-received 2D and 3D C/C composites exhibit manufacturing defects and porosity attributed to the graphitization process and the dissimilar expansion properties between the matrix and the fibers. Thermal cycle annealing has shown to remove most of the manufacturing defects, leading to a stable structure. Based on the experimental findings, thermal annealing of 2D and 3D C/C composites to be used as targets or collimators is recommended to remove manufacturing defects, thus making the materials thermally stable. (iii) The newly developed MoGRCF compound grade under consideration for beam collimation in the high luminosity LHC and in its as-received, uni-irradiated state is characterized by grain inhomogeneity and the presence of intermediate Mo carbide phases, all attributed to the manufacturing process. This appears to be a concern regarding the response of the compound in beam. (iv) Modest irradiation levels of $\sim 6 \times 10^{18}$ have shown in this study to induce significant changes in the microstructure of this MoGRCF. These changes also manifested themselves as structural degradation, a finding that puts into serious question the use of this particular grade of the compound in the LHC collimation system. X-ray diffraction evaluation at these modest fluences confirmed the microstructural instability of the compound. (v) Irradiation exposure to a fluence $\geq 2 \times 10^{20}$ p/cm$^2$ has resulted in the complete failure of the compound despite the fact that irradiation temperatures were in excess of 400°C that should have triggered irradiation damage annealing. Based on the high and low fluence response of this MoGRCF compound grade, its use in the high luminosity LHC is discouraged. Improved grades are currently under study. (vi) 2D-C/C composite used in the phase-I LHC collimator structures and the 3D-C/C composite under consideration for pion production targets in the megawatt-level Neutrino Factory and Long Baseline Neutrino Facility accelerator have been assessed to have a fluence threshold of $\sim 5 \times 10^{20}$ p/cm$^2$, above which significant changes have been shown to occur in their lattice structure of the order of 3% increase in the lattice parameter that indicates radiation-induced growth or density reduction. This effect that may be taken into consideration when establishing the useful lifetime of a high-power target (or collimator) composed of these materials by establishing it as a low bound limit. (vii) The study revealed the similarity between the effects of fast neutrons and of energetic protons on the microstructure of carbon-based materials (i.e., graphite and C/C composites) and, in particular, around the threshold fluence of $\sim 5 \times 10^{20}$ cm$^{-2}$, an important finding of this study which further supports the establishment of operational limits based on the results reported. (viii) It was experimentally verified that the 2D- and 3D-C/C composites achieve dimensional stability through annealing by removing the preexisting porosity and defects. Preannealing to temperatures $\geq T_{\text{irrad}}$ anticipated during the specific application is recommended for these material structures.

## Future work

Prompted by the observed premature structural damage of the irradiated MoGRCF grade (21.51% Mo and 78.485% carbon of atomic fraction), irradiation experiments of two new MoGRCF grades characterized by a lower Mo content and the addition of Ti in the matrix have been launched at BLIP to assess the effect of the composition of the compound on the irradiation damage resistance. In the same experiments, a 2D C/C (Toyo Tanso AC-150) composite is being irradiated, aiming to reach a peak fluence of $\sim 10^{21}$ p/cm$^2$. The results of this study, including postirradiation analysis, will be presented in an upcoming paper. In addition, irradiation experiments utilizing spallation-induced fast neutrons with the objective to compare the effects of energetic protons with those of fast neutrons and establish damage in terms of displacements-per-atom and property evolution correlation have been initiated.

## ACKNOWLEDGMENTS

This research used resources of the Center for Functional Nanomaterials, which is a U.S. DOE Office of Science Facility, at Brookhaven National Laboratory under Contract No. DE-SC-0012704. This research used resources at X17B1 beamline of the National Synchrotron Light Source, a U.S. Department of Energy (DOE) Office of Science User Facility operated for the DOE Office of Science by Brookhaven National Laboratory under Contract No. DE-AC02-98CH10886.